\documentclass[prl,twocolumn,superscriptaddress,showpacs]{revtex4}

\usepackage{graphicx}
\usepackage{amssymb,amsmath,color}
\usepackage{algorithm2e}

\newcommand{\ket}[1]{|#1\rangle}

\begin{document}

\title{Realization of a scalable Shor algorithm}

\date{\today}

\author{T. Monz}
\affiliation{Institut f\"ur Experimentalphysik, Universit\"at
Innsbruck, Technikerstr. 25, A-6020 Innsbruck, Austria}

\author{D. Nigg}
\affiliation{Institut f\"ur Experimentalphysik, Universit\"at
Innsbruck, Technikerstr. 25, A-6020 Innsbruck, Austria}

\author{E. A. Martinez}
\affiliation{Institut f\"ur Experimentalphysik, Universit\"at
Innsbruck, Technikerstr. 25, A-6020 Innsbruck, Austria}

\author{M. F. Brandl}
\affiliation{Institut f\"ur Experimentalphysik, Universit\"at
Innsbruck, Technikerstr. 25, A-6020 Innsbruck, Austria}

\author{P. Schindler}
\affiliation{Institut f\"ur Experimentalphysik, Universit\"at
Innsbruck, Technikerstr. 25, A-6020 Innsbruck, Austria}

\author{R. Rines}
\affiliation{Center for Ultracold Atoms, Massachusetts Institute of Technology,
77 Massachusetts Ave, Cambridge, MA, USA}

\author{S. X. Wang}
\affiliation{Center for Ultracold Atoms, Massachusetts Institute of Technology,
77 Massachusetts Ave, Cambridge, MA, USA}

\author{I. L. Chuang}
\affiliation{Center for Ultracold Atoms, Massachusetts Institute of Technology,
77 Massachusetts Ave, Cambridge, MA, USA}

\author{R.~Blatt}
\affiliation{Institut f\"ur Experimentalphysik, Universit\"at
Innsbruck, Technikerstr. 25, A-6020 Innsbruck, Austria}
\affiliation{Institut f\"ur Quantenoptik und Quanteninformation,
\"Osterreichische Akademie der Wissenschaften, Otto-Hittmair-Platz
1, A-6020 Innsbruck, Austria}


\begin{abstract}
  Quantum computers are able to outperform classical algorithms. This
  was long recognized by the visionary Richard Feynman who pointed out
  in the 1980s that quantum mechanical problems were better solved
  with quantum machines. It was only in 1994 that Peter Shor came up
  with an algorithm that is able to calculate the prime factors of a
  large number vastly more efficiently than known possible with a
  classical computer \cite{Shor}. This paradigmatic algorithm
  stimulated the flourishing research in quantum information
  processing and the quest for an actual implementation of a quantum
  computer. Over the last fifteen years, using skillful optimizations,
  several instances of a Shor algorithm have been implemented on
  various platforms and clearly proved the feasibility of quantum
  factoring \cite{citeulike:5725227, citeulike:11534452,
    citeulike:11389323, citeulike:3430059, citeulike:2365123,
    shor_nmr_chuang}. For general scalability, though, a different
  approach has to be pursued \cite{fake_shor_smolin}. Here, we report
  the realization of a fully scalable Shor algorithm as proposed by
  Kitaev \cite{iter_shor_kitaev}. For this, we demonstrate factoring
  the number fifteen by effectively employing and controlling seven
  qubits and four ``cache-qubits'', together with the implementation
  of generalized arithmetic operations, known as modular multipliers.
  The scalable algorithm has been realized with an ion-trap quantum
  computer exhibiting success probabilities in excess of 90\%.
\end{abstract}

\maketitle

Shor's algorithm for factoring integers~\cite{Shor} is one of the
examples where a quantum computer (QC) outperforms the most efficient
known classical algorithms. Experimentally, its implementation is
highly demanding as it requires both a sufficiently large quantum
register and high-fidelity control. Clearly, such challenging
requirements raise the question whether optimizations and experimental
shortcuts are possible. While optimizations, especially
system-specific or architectural, certainly are possible, for a
demonstration of Shor's algorithm to be \emph{scalable} special care
has to be taken not to oversimplify the implementation - for instance
by employing knowledge about the solution prior to the actual
experimental implementation - as pointed out in
Ref.~\citenum{fake_shor_smolin}.

In order to elucidate the general task at hand, we first explain
and exemplify Shor's algorithm for factoring the number 15 in a
(quantum) circuit model. Subsequently, we show how this circuit model
is translated for and implemented with an ion-trap quantum computer.

How does Shor's algorithm work? Here is a classical recipe to find the
factors of a large number. As an example, assume the number we want to
factor is $N=15$. Then pick a random number $a \in [2,N-1]$ (which we
call the \emph{base} in the following), say $a=7$. Check if the
greatest common divisor gcd$(a,N) = 1$, otherwise a factor is already
determined. This is the case for $a=\{3,5,6,9,10,12\}$. Next,
calculate the modular exponentiations $a^x\mod N$ for $x = 0,1,2...$
and find its period $r$: the first $x>0$ such that $a^x\mod N=1$.
Given the period $r$, finding the factors requires calculating the
greatest common divisors of $a^{r/2}\pm 1$ and $N$, which is
classically efficiently possible - for instance using Euclid's
algorithm. For our example ($N=15, a=7$) the modular exponentiation
yields {1, 7, 4, 13, 1, ...}, which has period 4. The greatest common
divisor of $a^{r/2}\pm 1=7^{4/2}\pm 1=\{48,50\}$ and $N=15$ are
$\{3,5\}$, the non-trivial factors of $N$. For the chosen example
$N=15$, the cases $a=\{4,11,14\}$ have periodicity $r=2$ and would
only require a single multiplication step ($a^2 \mod N = 1$), which is
considered an ``easy'' case~\cite{fake_shor_smolin}. Note that the
periodicity for a chosen $a$ can not be predicted.

How can this recipe be implemented in a QC? A QC also has to calculate
$a^x \mod N$ in a computational register for $x=0,1,2...$ and then
extract $r$. However, using the quantum Fourier-transform (QFT), this
can be done with high probability in a single step (compared to $r$
steps classically). Here, $x$ is stored in a quantum
register consisting of $k$ qubits, or period-register, which is in a superposition of 0 to $2^k-1$. The superposition in the period-register on its own does not
provide a speedup compared to a classical computer. Measuring the
period-register would collapse the state and only return a single
value, say $x_1$, and the corresponding answer to $a^{x_1} \mod N$ in
the computational register. However, if the QFT is applied to the
period-register, the period of $a^x \mod N$ can be extracted from
$\mathcal{O}(1)$ measurements.

\begin{figure*}
 \begin{center}
   \includegraphics[width=2\columnwidth]{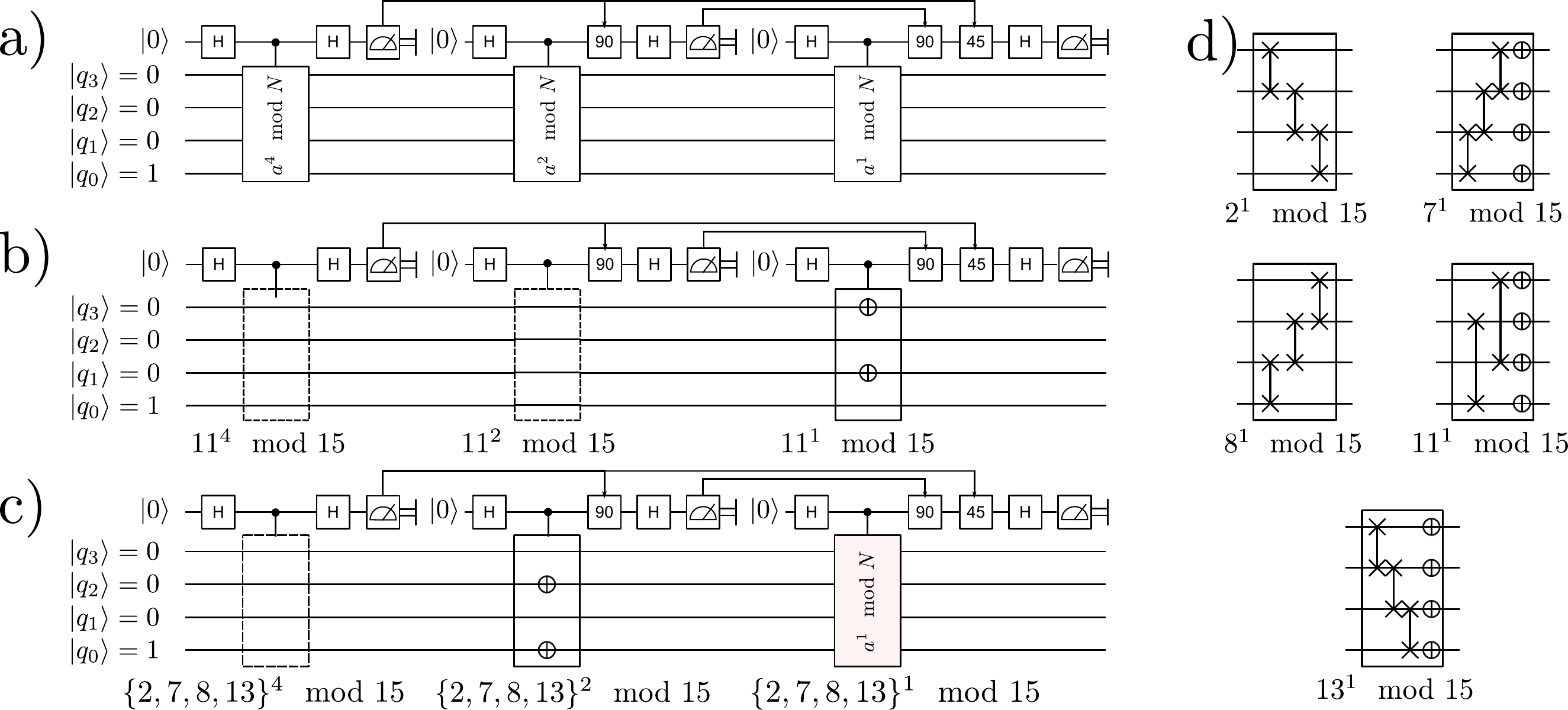}
   \caption{Circuit diagram of Shor's algorithm for factoring 15 based
     on Kitaev's approach for: a) a generic base $a$; and the specific
     circuit representations for the modular multipliers; b) The
     actual implementation for factoring 15 to base 11, optimised for
     the single input state it is subject to; c) Kitaev's approach to
     Shor's algorithm for the bases \{2,7,8,13\}: the optimised map of
     the first multiplier is identical in all 4 cases, the last
     multiplier is implemented with full modular multipliers as
     depicted in d); d) Circuit diagrams of the modular multipliers of
     the form $a \mod N$ for bases a=\{2,7,8,11,13\}.}
  \label{fig.shor_shem}
 \end{center}
\end{figure*}

What are the requirements and challenges to implement Shor's
algorithm? First, we focus on the period-register, to subsequently
address modular exponentiation in the computational register.
Factoring $N$, an $n=\lceil \log_2(N) \rceil$-bit number requires a
minimum of $n$ qubits in the computational register (to store the
results of $a^x \mod N$) and generally about $2n$ qubits in the
period-register~\cite{mike_n_ike}. Thus even a seemingly simple
example such as factoring 15 (an $n=4$ -bit number), would require
$3n=12$ qubits when implemented in this straightforward way. These
qubits then would have to be manipulated with high fidelity gates.
Given the current state-of-the-art control over quantum
systems~\cite{summaryarticle}, such an approach likely yields an
unsatisfying performance. However, a full quantum implementation of
this part of the algorithm is not really necessary. In
Ref.~\citenum{iter_shor_kitaev} Kitaev notes that, if only the
classical information of the QFT (such as the period $r$) is of
interest, $2n$ qubits subject to a QFT can be replaced by a single
qubit. This approach, however, requires qubit-recycling (specifically:
in-sequence single-qubit readout and state reinitialization) paired
with feed-forward to compensate for the reduced system size.

In the following, Kitaev's QFT will be referred to as KQFT$^{(M)}$. It
replaces a QFT acting on $M$ qubits with a semiclassical QFT acting
repeatedly on a single qubit. Similar applications of Kitaev's
approach to a semiclassical QFT in quantum algorithms have been
investigated in
Refs.~\cite{shor_iter_griffiths,shor_iter_plenio,shor_iter_ekert}. For
the implementation of Shor's algorithm, Kitaev's approach provides a
reduction from the previous $n$ computational qubits and $2n$ QFT
qubits (in total $3 n$ qubits) to only $n$ computational-qubits and 1
KQFT$^{(2n)}$ qubit (in total $n+1$ qubits).

\begin{figure*}
  \includegraphics[width=2\columnwidth]{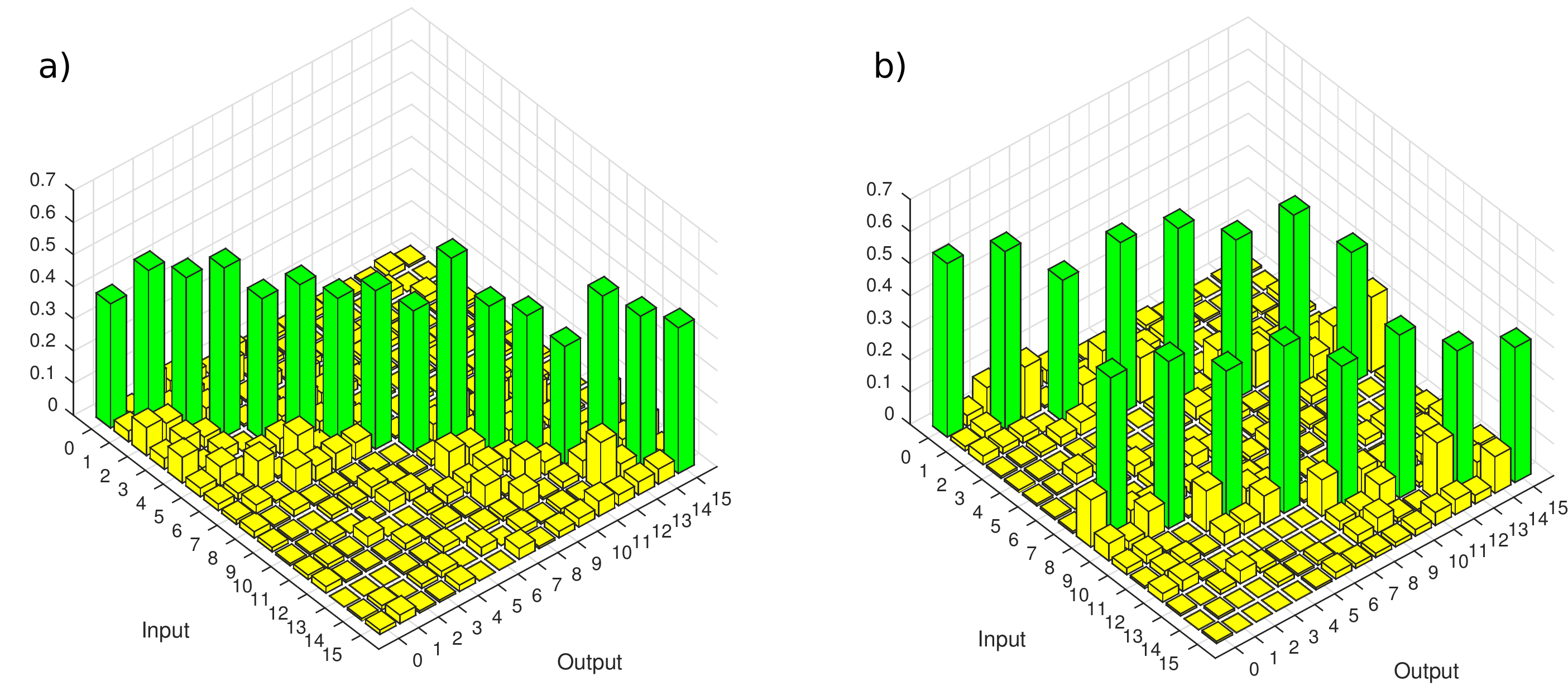}
  \caption{Experimentally obtained truth table of the controlled 2
    modular 15 multiplier: a) with the control-qubit being in state 0,
    the truth table corresponds to the identity operation; b) when the
    control qubit triggers the multiplication, the truth table
    illustrates the multiplication of the input state with 2 modular
    15.}
  \label{fig.mod_mult_2}
\end{figure*}

A notably more challenging aspect than the QFT, and the second
key-ingredient of Shor's algorithm, is the modular exponentiation,
which admits these general simplifications:

(i) Considering Kitaev's approach (see Fig.~\ref{fig.shor_shem}), the
input state $\ket{1}$ (in decimal representation) is subject to a
conditional multiplication based on the most-significant bit $k$ of
the period register. At most there will be two results after this
first step. It follows that, for the very first step it is sufficient
to implement an optimized operation that conditionally maps
$\ket{1}\rightarrow \ket{a^{2^k} \mod N}$. Considering the importance
of a high-fidelity multiplication (with its performance being
fed-forward to all subsequent qubits), this efficient simplification
improves the overall performance of experimental realizations.

(ii) Subsequent multipliers can similarly be replaced with maps by
considering only possible outputs of the previous multiplications.
However, using such maps will become exponentially more challenging,
as the number of input and output states to be considered grows
exponentially with the number of steps: after $n$ steps, $2^n > N$
possible outcomes need to be considered - a numerical task as
challenging as factoring $N$ by classical means. Thus, controlled full
modular multipliers need to be implemented. Fig.~\ref{fig.mod_mult_2}
shows the experimentally obtained truth table for the modular
multiplier $(2\mod 15)$ (see also supplementary material for modular
multipliers with bases $\{7,8,11,13\}$). These quantum circuits can be
efficiently derived from classical procedures using a variety of
standard techniques for reversible quantum arithmetic and local logic
optimization~\cite{efficient_multipliers_1,efficient_multipliers_2}.

(iii) The very last multiplier allows one more simplification:
Considering that the actual results of the modular exponentiation are
not required for Shor's algorithm (as only the period encoded in the
period-register is of interest), the last multiplier only has to
create the correct amount of correlations between the period register
and the computation register. Local operations after the conditional
(entangling) operations may be discarded to facilitate the final
multiplication without affecting the results of the implementation.

(iv) In rare cases, certain qubits are not subject to operations in
the computation. Thus, these qubits can be removed from the algorithm
entirely.

For larger scale quantum computation, optimization steps (i), (iii) and
(iv) will only marginally effect the performance of the implementation.
They represent only a small subset of the entire computation which
mainly consists of the full modular multipliers. Thus, the realization
of these modular multipliers is a core requirement for scalable
implementations of Shor's algorithm.

Furthermore, Kitaev's approach requires in-sequence measurements,
qubit-recycling to reset the measured qubit, feed-forward of gate
settings based on previous measurement results, as well as numerous
controlled quantum operations - tasks that have not been realized in a
combined experiment so far.

We demonstrate these techniques in our realization of Shor's algorithm
in an ion-trap quantum computer, with five $^{40}$Ca$^+$ ions in a
linear Paul trap. The qubit is encoded in the ground state
$S_{1/2}(m=-1/2)=\ket{1}$ and the metastable state
$D_{5/2}(m=-1/2)=\ket{0}$. The universal set of quantum gates consists
of the entangling M{\o}lmer-S{\o}renson interaction~\cite{ms_gate},
collective operations of the form $\exp(-i \frac{\theta}{2} S_\phi)$
with $S_\phi = \sum_i \sigma_\phi^{(i)}$, $\sigma_\phi^{(i)} =
\cos(\phi)\sigma_x^{(i)} + \sin(\phi)\sigma_y^{(i)}$,
$\sigma_{\{x,y\}}^{(i)}$ the Pauli operators of qubit $i$,
$\theta=\Omega t$ determined by the Rabi frequency $\Omega$ and laser
pulse duration $t$, $\phi$ determined by the relative phase between
qubit and laser, and single qubit phase rotations induced by localized
AC-Stark shifts (for more details see the supplementary material and
Ref.~\citenum{order_finding_phips}). Unitary operations illustrated in
Fig.~\ref{fig.shor_shem} are decomposed into primitive components such
as two-target C-NOT and C-SWAP gates (or gates with global symmetries
such as the four-target C-NOT employed here), from which an adaptation
of the GRAPE algorithm~\cite{optimal_control} can efficiently derive
an equivalent sequence of laser pulses acting on only the relevant
qubits. The problem with this approach is that the resulting sequence
generally includes operations acting on all qubits. Implementing the
optimized 3-qubit operations on a 5-ion string therefore requires
decoupling of the remaining qubits from the computation space. We
spectroscopically decouple qubits by transferring any information from
$\ket{S}\rightarrow \ket{D'}=D_{5/2}(m=-5/2)$ and $\ket{D}\rightarrow
\ket{S'}=S_{1/2}(m=1/2)$. Here, the subspace $\{\ket{S'},\ket{D'}\}$
serves as a readily available ``quantum cache'' to store and retrieve
quantum information in order to facilitate quantum computations.

Finally, to complete the toolbox necessary for a Kitaev's approach to
Shor's algorithm, we also implement single qubit readout (by encoding
all other qubits in the $\{\ket{D},\ket{D'}\}$ subspace and subsequent
electron shelving~\cite{electron_shelving_proposal} on the
$S_{1/2}\leftrightarrow P_{1/2}$ transition), feed-forward (by storing
counts detected during the single-qubit readout~\cite{teleport_riebe}
in a classical register and subsequent conditional laser pulses) and
state-reinitialization (using optical pumping for the ion, and
Raman-cooling~\cite{wineland_bible,raman_cooling_zoller} for the
motional state of the ion string). The pulse sequences and additional
information on the implementation on the modular multipliers are
available as supplementary material.

The key differences of our implementation with respect to previous
realizations of Shor's algorithm are: a) the entire quantum register
is employed, without sparing qubits that don't partake in the
calculation; b) besides the trivial first multiplication step
(corresponding to $r=2$ for $a=\{4,11,14\}$, realized only once for
$a=11$), all non-trivial modular multipliers $a=\{2,7,8,13\}$ have
been realized and applied; and c) Kitaev's originally proposed scheme
is implemented with complete qubit recycling -- doing both readout and
reinitialization on the very same physical qubit. This is especially
important for factoring 15 with base \{2,7,8,13\}, as at least two
steps are required for the semiclassical QFT. In our realization we go
beyond the minimal implementation of Shor's algorithm and not only
employ all 7 qubits (comprised of 4 physical qubits in the
computational register, 1 qubit in the periodicity register - recycled
twice, plus additional cache qubits), but also include multiplication
with up to the fourth power (although they correspond to the identity
operation). This represents a realistic attempt at a scalable
implementation of Shor's algorithm as the entire qubit register
remains subject to decoherence processes along the computation, and no
simplifications are employed which presume prior knowledge of the
solution.

\begin{figure}
  \includegraphics[width=\columnwidth]{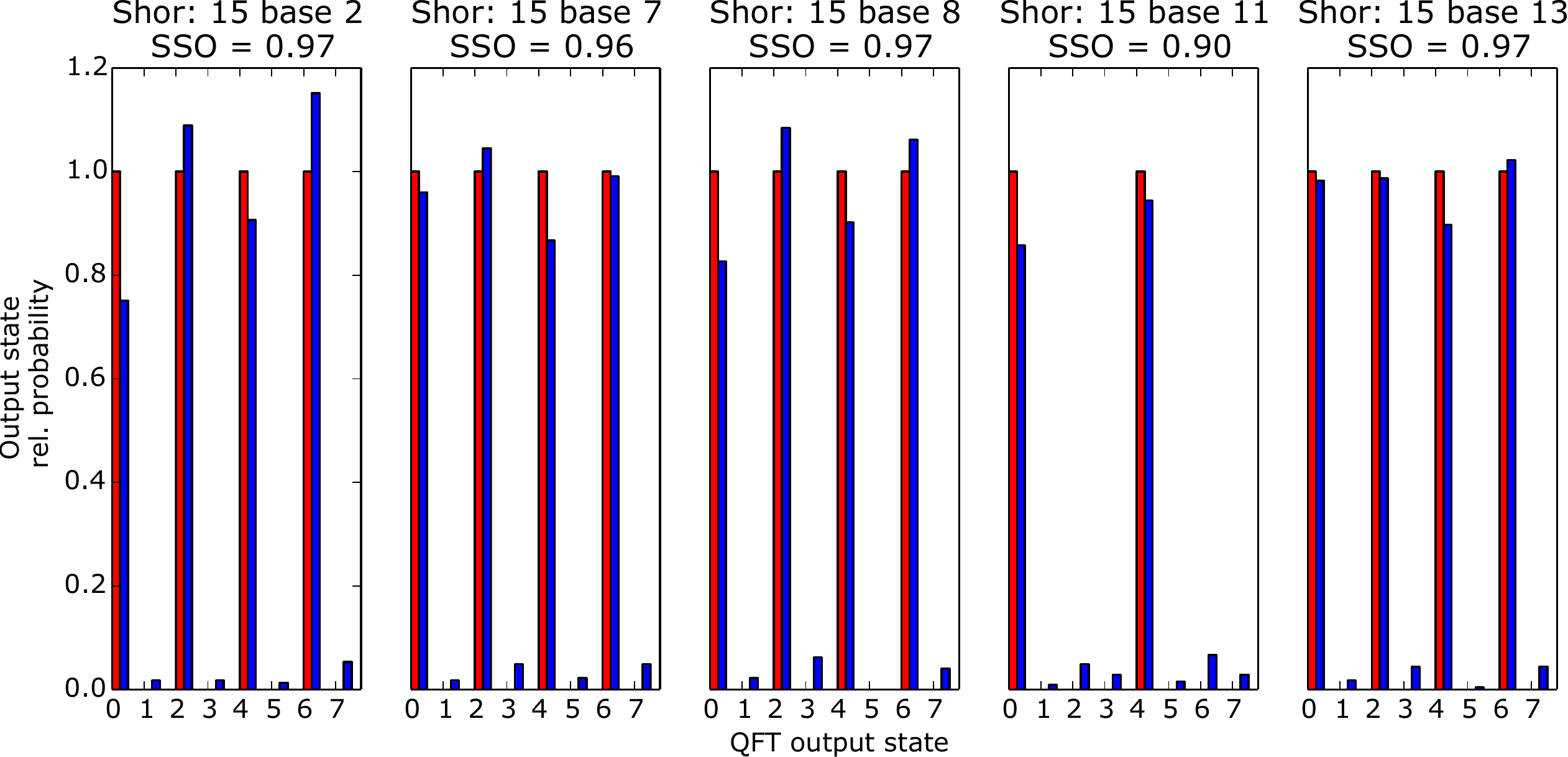}
  \caption{Results and correct order-assign probability for the
    different implementations to factor 15: a) 3-digit results (in
    decimal representation) of Shor's algorithm for the different
    bases. The ideal data (red) for periodicity $\{2,4\}$ is
    superimposed on the raw data (blue). The squared statistical
    overlap is larger than 90\% for all cases.}
  \label{fig.shor_results}
\end{figure}

The measurement results for base $a=\{2,7,8,11,13\}$ with
periodicities $r=\{4,4,4,2,4\}$ are shown in
Fig.~\ref{fig.shor_results}. In order to quantify the performance of
the implementation, previous realizations mainly focused on the
squared statistical overlap (SSO)~\cite{chiaverini_semiclas_fourier},
the classical equivalent to the Uhlmann fidelity~\cite{mike_n_ike}.
While we achieved an SSO of \{0.968(1), 0.964(1), 0.966(1), 0.901(1),
0.972(1)\} for the case of a=\{2,7,8,11,13\}, we argue that this does
not answer the question of a user in front of the quantum computer:
``What is the periodicity?'' Shor's algorithm allows one to deduce the
periodicity with high probability from a single-shot measurement,
since the output of the QFT is, in the exact case, a ratio of
integers, where the denominator gives the desired periodicity. This
periodicity is extracted using a continued fraction expansion, applied
to $x/2^k$, a good approximation of the ideal case when $k$, the
number of qubits, is sufficiently large. For the realised examples,
the probabilistic nature of Shor's algorithm becomes clear: the output
state $0$ never yields any information. For periodicity $4$ (and 3
qubits in the period-register), the output state $4$ suggests a
fraction $\frac{4}{2^3}=\frac{1}{2}$, thus a periodicity of $2$ and
also fails. For peridocity $4$, only the output states $2$ and $6$
allow one to deduce the correct periodicity. In our realisations to
bases $a=\{2,7,8,11,13\}$, the probabilities to obtain output states
that allow the derivation of the correct periodicity are
$\{56(2),51(2),54(2), 47(2), 50(2)\}$\%. Thus, a confidence that the
correct periodicity is obtained at a level of more than 99\%, requires
the experiment to run about 8 times.

In summary, we have presented the realization of Kitaev's vision
to realize a scalable Shor's algorithm with 3-digit resolution to
factor 15 using bases \{2,7,8,11,13\}. Here, a semiclassical QFT
combined with single-qubit readout, feed-forward and qubit recycling
was successfully employed. Compared to the traditional algorithm, the
required number of qubits can thus be reduced by almost a factor of 3.
Furthermore, the entire quantum register has been subject to the
computation in a ``black-box'' fashion. Employing the equivalent of a
quantum cache by spectroscopic decoupling significantly facilitated
the derivation of the necessary pulse sequences to achieve
high-fidelity results. In the future, spectroscopic decoupling might
be replaced by physically moving the qubits from the computational
zone using segmented traps~\cite{scale_iontraps}.

Our investigations also reveal some open questions and problems for
current and upcoming realizations of Shor's algorithm, which also
apply to several other large-scale quantum algorithms of interest:
particularily, finding system-specific implementations of suitable
pulse sequences to realize the desired evolution. The presented
operations were efficiently constructed from classical circuits, and
decomposed into manageable unitary building blocks (quantum gates) for
which pulse sequences were obtained by an adapted GRAPE algorithm.
Thus, the presented successful implementation in an ion-trap quantum
computer demonstrates a viable approach to a scalable Shor algorithm.

We gratefully acknowledge support by the Austrian Science Fund (FWF),
through the SFB FoQus (FWF Project No. F4002-N16), by the European
Commission (AQUTE), the NSF iQuISE IGERT, as well as the Institut
f\"ur Quantenoptik und Quanteninformation GmbH. EM is a recipient of a
DOC Fellowship of the Austrian Academy of Sciences. This research was
funded by the Office of the Director of National Intelligence (ODNI),
Intelligence Advanced Research Projects Activity (IARPA), through the
Army Research Office grant W911NF-10-1-0284. All statements of fact,
opinion or conclusions contained herein are those of the authors and
should not be construed as representing the official views or policies
of IARPA, the ODNI, or the U.S. Government.

\bibliography{shor_bib}

\appendix

\newpage

\section{Supplementary Material}

\subsection{Pulse sequences}

In the following, the pulse sequences employed in the experiment are discussed in more detail. The nomenclature is as follows: The
collective operations on the $S_{1/2}(m=-1/2)\leftrightarrow
D_{5/2}(m=-1/2)$ transitions, addressing all ion-qubits, realize the
unitary operation $$ R(\theta, \phi) = \exp(-i \frac{\pi}{2} \theta
S_{\phi})$$ with the collective spin operator $$S_{\phi}= \sum_i
\sigma_\phi^{(i)} = \sum_i \cos(\phi \pi) \sigma_x^{(i)} + \sin(\phi
\pi) \sigma_y^{(i)}$$ based on the Pauli operators
$\sigma_{\{x,y,z\}}^{(i)}$ acting on qubit qubit $i$. Here, the
rotation angle $\theta$ is defined by $\theta=\frac{\Omega t}{\pi}$
with the Rabi frequency $\Omega$ and the laser pulse duration $t$. In
this notation, a bit flip around $\sigma_x$ corresponds to $R(1,0)$.
The collective operations are supplemented by single-qubit phase
shifts of the form $$S_z(\theta,i) = \exp(-i \frac{\theta \pi}{2}
\sigma_z^{(i)}).$$ The phase shift is realized by illuminating a
single qubit with a tightly focused laser beam detuned $-20$~MHz from
the carrier transition. Here, the induced AC-Stark shift $\Delta_{AC}$
implements the desired phase shift, with the rotation angle $\theta=
\frac{\Delta_{AC} t}{\pi}$ depending on the pulse duration $t$. In
combination, collective operations and single-qubit phase shifts allow
us to implement arbitrary local operations. A universal set of quantum
gates, capable of implementing any desired unitary operation, can
be realized by combining these arbitrary local operations with an
entangling interaction. In our experiment, we employ the
M{\o}lmer-S{\o}rensen (MS) interaction~\cite{ms_gate} to realize
entangling operations of the form $$MS(\theta)=\exp(-i \frac{\pi}{4}
\theta S_{x}^2)$$ with $S_{x}=\sum \sigma_x^{(i)}$. Using this
notation, the maximally entangling $MS(\frac{1}{2})$ operation
applied onto the $N$-qubit state $\ket{0\ldots0}$ directly creates the
$N$-qubit GHZ state.

\subsection{Single-qubit measurement}

Electron-shelving~\cite{electron_shelving_proposal} on the $S_{1/2}
\leftrightarrow P_{1/2}$ transition addresses, and thus projects, all
qubits of the quantum register. For Kitaev's implementation, however,
only one qubit needs to be measured. With collective illumination,
this can be achieved by transfering quantum information
encoded in qubits that should not be measured into the $D$-state
manifold. Here, the quantum information is protected against shelving
light on the $S_{1/2} \leftrightarrow P_{1/2}$ transition - the ion
will not scatter any photons. Using light resonant with the
$S_{1/2}(m=-1/2) \leftrightarrow D_{5/2}(m=-5/2)$ transition (denoted
by $R_2(\theta,\phi)$), a refocusing sequence of the form $R_2(0.5,0)
\cdot S_z(1,i) \cdot R_2(0.5,0)$ 
efficiently encodes all but qubit $i$ in $D_{5/2}(m=-1/2)$ and
$D_{5/2}(m=-5/2)$. Subsequently, the entire quantum register may be
subject to shelving light, yet only qubit $i$ will be projected.

\subsection{In-sequence detection and feed-forward}

\begin{figure}[h]
  \centering
  \includegraphics[width=\columnwidth]{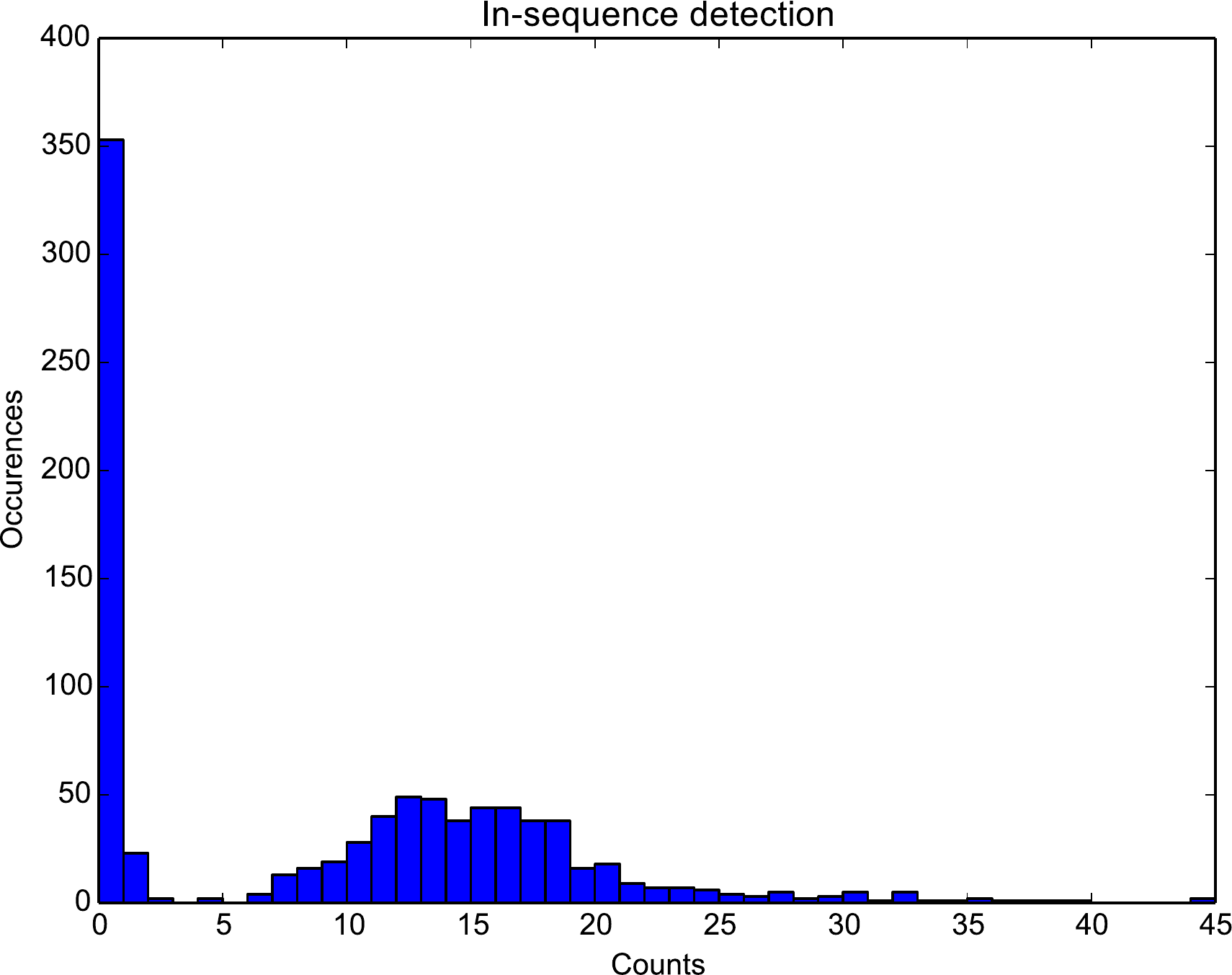}
  \caption{In-sequence photon-count histogram: Using a detection
    window of 300 $\mu s$, the photomultiplier tube collects on
    average 0.07~{counts} when the qubit is in state $D$ and
    14.4~{counts} when it is in state $S$. As can be seen in the
    figure, these two Poisson distributions are well distinguishable.}
  \label{fig:det_hist}
\end{figure}

When all qubits that need to be protected against projection have been
encoded in the $\{D_{5/2}(m=-1/2),D_{5/2}(m=-5/2)\}$ manifold, light
at 397~nm resonant with the $S_{1/2}\leftrightarrow P_{1/2}$
transition state-dependently scatters photons an the remaining
ion-qubits. The illumination time is set to 300~$\mu$s. A histogram
of the photon counts detected at the photomultiplier tube is shown in
Fig.~\ref{fig:det_hist}. Using counter electronics with discriminator
set at 4 counts within the detection window, the state $D$ with a mean
count rate of 0.24~{counts/ms} (or 0.07~counts within the detection
window) and state $S$ with a mean countrate of 48~{counts/ms} (or
14.4~counts in the detection window) can be distinguished with a
confidence better than 99.8\%. The boolean output of the discriminator
is subsequently used in the electronics for state-dependent pulses and
thus state-dependent operations.

\subsection{Recooling and Qubit-reset}

Scattering photons during the detection window heats the ion-string
and can lower the quality of subsequent quantum operations applied
to the register. Therefore recooling of the ion-string after the
illumination with electron-shelving light is necessary. However, this
recooling must not destroy any quantum information stored in the other
qubits. Considering that the hidden quantum information is stored in
the $D_{5/2}$ manifold, we employ 3-beam
Raman-cooling~\cite{wineland_bible,raman_cooling_zoller} in the
$S_{1/2}\leftrightarrow P_{1/2}$ manifold. The Raman light field,
consisting of $\sigma^+$ and $\pi$ light with respect to the
quantization axis, is detuned by 1.5~GHz from the resonant $S_{1/2}
\leftrightarrow P_{1/2}$ transition. The relative detuning between
$\sigma^+$ and $\pi$ is chosen such that it creates resonant coupling
between $S_{1/2}(m=-1/2) \otimes \ket{n} \leftrightarrow
S_{1/2}(m=1/2) \otimes \ket{n-1}$, with $\ket{n}$ representing the
quantized axial state of motion of the ion. The transfer is reset by
resonant $\sigma^-$ light. Raman cooling is employed for 500 $\mu$s.
The qubit is reinitialized after cooling by an additional 50~$\mu$s
of $\sigma^-$ light. However, if the measured qubit was found to be in
state $D$, neither does the measurement heat the ion string nor does
the Raman cooling affect the register. Therefore the qubit is
transferred from $D_{5/2}(m=-1/2)$ to $S_{1/2}(m=1/2)$ (which was
depleted by the previous 50~$\mu$s of $\sigma^-$). An additional
pulse of $\sigma^-$ light for 50~$\mu$s finally initializes the
qubit, regardless whether it was projected into $S$ or $D$. During the
entire time when the qubit is subject to Raman cooling or initializing
$\sigma^-$ light, a repump laser at 866~nm is applied to prevent
population trapping in the $D_{3/2}$ manifold due to spontaneous decay
from the $P_{1/2}$ state to $D_{3/2}$.

\subsection{Pulse sequence optimisation}

For a sufficiently large Hilbert-space it will no longer be possible
to directly optimize unitary operations acting on the entire register.
Decomposing the necessary unitary operations into building blocks
acting on smaller register sizes will allow one the use of optimized
pulse sequences for large-scale quantum computation. From a
methodological point of view it may be preferred to physically decouple
the qubits from any interactions (for instance by splitting and moving
part of ion-qubit quantum register out of an interaction region, such
as proposed in Ref.~\citenum{scale_iontraps}). However, given the
technical requirements and challenges for splitting and moving
ion-strings, we focus on spectroscopically decoupling certain
ion-qubits from the interaction. In particular, we spectroscopically
decouple an ion from subsequent interaction by transferring any
quantum information from the $\{S_{1/2}(m=-1/2),D_{5/2}(m=-1/2)\}$
manifold to the $\{S_{1/2}(m=1/2),D_{5/2}(m=-5/2)\}$ manifold using
refocusing techniques on the $D_{5/2}(m=-1/2)\leftrightarrow
S_{1/2}(m=1/2)$ and $S_{1/2}(m=-1/2)\leftrightarrow D_{5/2}(m=-5/2)$
transitions. Using this approach, we optimise the controlled swap
operation in a 3-qubit Hilbert space rather than a 5-qubit Hilbert
space.

\subsection{Controlled-SWAP}

The controlled-SWAP operation, also known as Fredkin operation, plays
a crucial role in the modular multiplication. For its implementation,
however, we could not derive a pulse sequence that can incorporate an
arbitrary number of spectator qubits --- qubits, that should be
subject to the identity operation --- in the presented case, i.e. 2
spectator qubits in the computational register. However, using
decoupling of spectator qubits, this additional requirement on the
implementation is not necessary. Using pulse sequence
optimization~\cite{optimal_control}, we obtained a sequence for the
exact three-qubit case as shown in Tab.~\ref{tab:fredkin}. In total
the sequence consists of 18 pulses, including 4 MS
interactions. 

\begin{table}[h!]
  \centering
  \begin{tabular}{|c|l||c|l|}
    \hline
    Pulse Nr. & Pulse & Pulse Nr. & Pulse \\
    \hline
    1 & $R(1/2,1/2)$ & 10 & $R(1/2,1)$ \\
    2 & $S_z(3/2,3)$ & 11 & $S_z(1/4,2)$\\
    3 & $MS(4/8)$     & 12 & $S_z(3/2,3)$\\
    4 & $S_z(3/2,2)$ & 13 & $MS(4/8)$\\
    5 & $S_z(1/2,3)$ & 14 & $S_z(3/2,2)$\\
    6 & $R(3/4,0)$  & 15 & $S_z(3/2,1)$\\
    7 & $MS(6/8)$     & 16 & $R(1/2,1)$\\
    8 & $S_z(3/2,2)$ & 17 & $S_z(3/2,1)$\\
    9 & $MS(4/8)$     & 18 & $S_z(3/2,2)$ \\
    \hline
  \end{tabular}
  \caption{Controlled SWAP operation: In a system of three ion-qubits, qubit 1 represents the control qubit and qubits $\{2,3\}$ are to be swapped depending on the state of the first qubit. Note that this sequence only works for three-qubit systems. Spectator qubits would not experience the identity operation.}
  \label{tab:fredkin}
\end{table}

\subsection{Four-Target Controlled-NOT}

The modular multipliers $(7\mod 15)$ and $(13\mod 15)$ require, besides
Fredkin operations, also CNOT operations acting on all qubits in the
computational register. Such an operation can be implemented (see
Ref.~\citenum{master_nebendahl} (p.90, eq. 5.21) ) with 2 MS operations
plus local operations only - regardless of the size of the
computational register. The respective sequence is shown in
Tab.~\ref{tab:ft-cnot}.

\begin{table}[h!]
  \centering
  \begin{tabular}{|c|l||c|l|}
    \hline
    Pulse Nr. & Pulse & Pulse Nr. & Pulse \\
    \hline
    1 & $R(1/2,1) $    & 6 & $MS(1/4) $\\
    2 & $S_z(3/2,1) $ & 7 & $R(3/4,0) $\\
    3 & $MS(3/4) $    & 8 & $S_z(3/2,1)$\\
    4 & $R(5/4,1)$   & 9 & $R(1/2,0)$\\
    5 & $S_z(1,1) $   & & \\
    \hline
  \end{tabular}
  \caption{Four-target controlled NOT: Depending on the state of qubit one, the remaining four qubits \{2-5\} are subject to a conditional NOT operation.}
  \label{tab:ft-cnot}
\end{table}

\subsection{Two-Target Controlled-NOT}

There exists an analytic solution to realize multi-target
controlled-NOT operations in the presence of spectator qubits with the
presented set of gates~\cite{master_nebendahl} - as required for the
$\{2,7,8,13\}^2 \mod 15$ multiplier. However, we find that performing
decoupling of subsets of qubits of the quantum register prior to the
application of the multi-target controlled-NOT operation presented
above both facilitates the optimisation, and improves the performance
of the realisation of a two-target controlled-NOT operation. Thus, the
required two-target controlled-NOT operation is implemented via (i)
decoupling qubits 2 and 4, (ii) performing a multi-target
controlled-not on all qubits with the first qubit acting as control,
and (iii) recoupling of qubits 2 and 4.


\subsection{Controlled Quantum Modular Multipliers}

Based on the decomposition shown in Fig.~\ref{fig.shor_shem}d) and the
respective pulse sequences outlined in the previous section, we
investigate the performance of the building blocks as well as the
respective conditional multipliers. In the following, the fidelities
are defined as mean probabilities and standard deviations to observe
the correct output state. The elements in the respective truth tables
have been obtained as average over 200 repetitions.
\begin{itemize}
\item The Fredkin operation, controlled by qubit 1 and acting on
  qubits $\{35,23,34,45\}$, yields fidelities of
  $\{76(4),73(6),72(4),68(7)\}\%$. These numbers are consistent with
  MS gate interactions at a fidelity of about 95\% acting on three
  ions (in the presence of two decoupled ions) and local operations at a
  fidelity of 99.3\%.
\item The 4-target CNOT gate operates at a fidelity of $86(3)\%$.
\item Considering the quality for modular multipliers of
  $(\{2,4,7,8,11,13\}\mod 15)$, we find fidelities of
  $\{48(5),40(5),50(6),46(5),38(5)\}\%$. This performance is
  consistent with the multiplication of the performance of the
  individual building blocks: $\{37(6), 36(5), 37(6), 48(5),
  36(5)\}\%$.
\end{itemize}

\begin{figure*}
  \includegraphics[height=0.9\textheight]{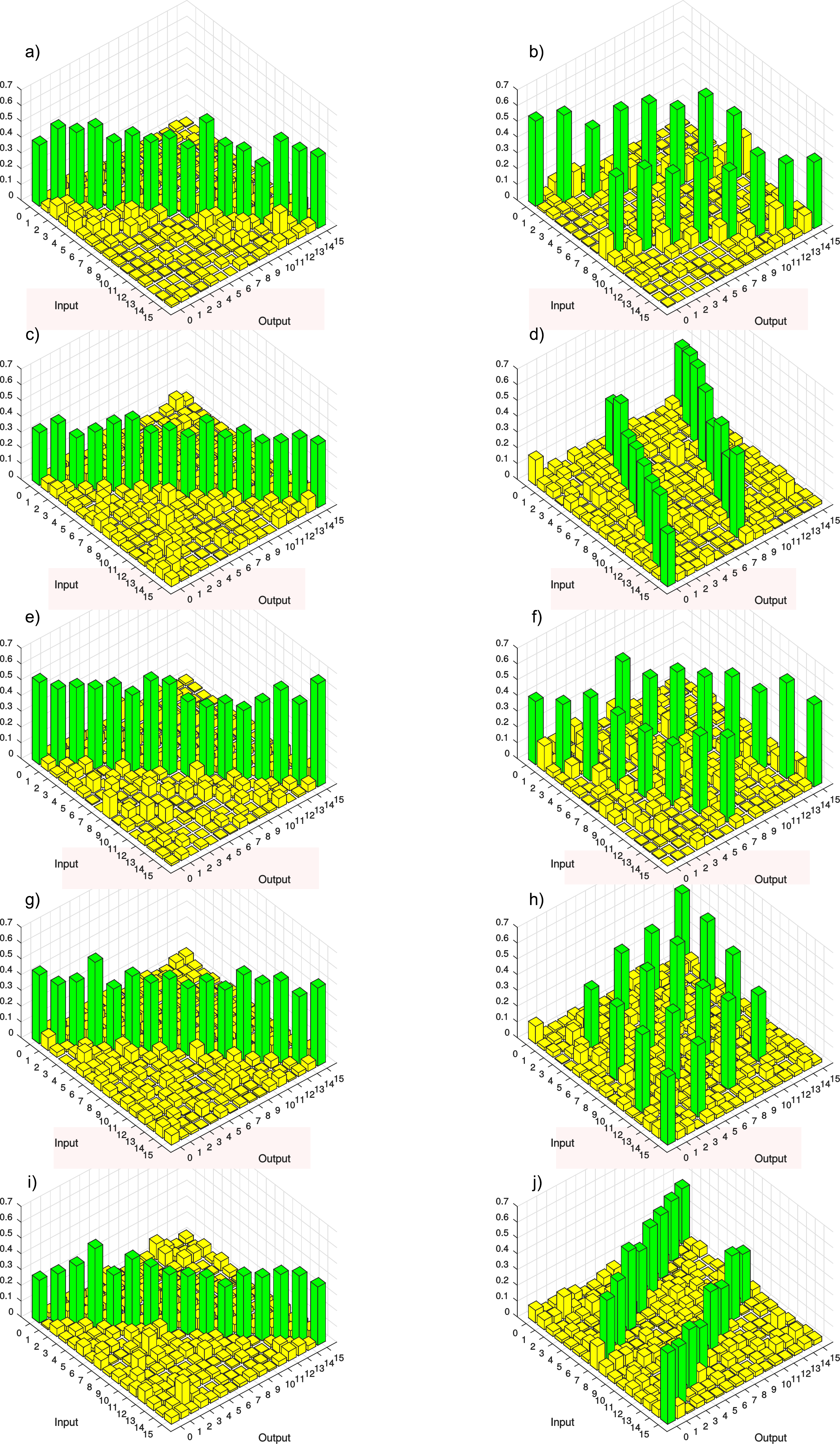}
  \caption{Controlled modular multipliers: While the full truth tables
    have been obtained, for improved visibility only the subset of
    data for the computational register (in decimal basis) is
    presented for modular multipliers $(\{2,7,8,11,13\}\mod 15)$ where
    the control bit maps from $\ket{0}$ to $\ket{0}$ (a,c,e,g,i) as
    well as when $\ket{1}$ maps onto $\ket{1}$ (b,d,f,h,j). When the
    control qubit is in state $\ket{0}$, one expects to find the
    identity operation implemented, as shown in (a,c,e,g,i). If the
    control qubit is in state $\ket{1}$, the input state gets
    multiplied by $(\{2,7,8,11,13\})\mod 15$. This behaviour is
    visually demonstrated as the output state increases in steps of
    $\{2,7,8,11,13\}$ until it reaches $15$, where the output is then
    returned to its value modulo 15.}
  \label{fig.mod_mult_all}
\end{figure*}

\end{document}